\documentstyle[12pt,gebmacros]{ioplppt} 

\newcommand{\mydate}{{\bf Preprint version as of April 3, 1997}, 
printed on \today}

\begin{document}
\jl{3}
\title{\begin{center}
{\large\bf Gutzwiller-correlated wave functions for degenerate bands:}\\
{\large\bf exact results in infinite dimensions}\end{center}}%
[Gutzwiller-correlated wave functions for degenerate bands]
\author{J B\"unemann\dag, F Gebhard\ddag\ and W Weber\dag}
\address{\dag\ 
Inst.~f.~Physik, Universit\"at Dortmund, D-44221 Dortmund, Germany}
\address{\ddag\
Inst.\ Laue--Langevin, B.~P.\ 156x, F-38042 Grenoble Cedex 9, France}
\begin{abstract}%
We introduce Gutzwiller-correlated wave functions for the
variational investigation of general multi-band Hubbard models. We set up
a diagrammatic formalism which allows us to evaluate analytically ground-state 
properties in the limit of infinite spatial dimensions.
In this limit recent results obtained within the Gutzwiller 
approximation are seen to become
exact for these wave functions. We further show that
the Slave Boson mean-field theory for degenerate bands
becomes variationally controlled at zero temperature in infinite dimensions.
Lastly, we briefly comment on the variational approach to the Anderson
transition in strongly correlated electron systems.
\end{abstract}

\maketitle

\section{Introduction}

During the last decades the theoretical investigation of strongly correlated
electron systems concentrated on the one-band Hubbard 
model~\cite{Hubbard}, which serves as the standard model for 
electrons with strong short-range interactions; for a recent review,
see~\cite{BUCH}. 
The Hubbard model was originally proposed for the description
of 3$d$~electrons in transition metals to explain
ferromagnetism and antiferromagnetism in 
iron and nickel and their oxides~\cite{Hubbard,Hubbard2}.
For these substances the band degeneracy and intra-atomic (Hund's rule)
exchange couplings apparently play an important role and,
consequently, multi-band Hubbard models need to be investigated.

Until recently, band degeneracies were considered mostly 
for immobile $f$~electrons which hybridize with featureless
conduction electrons.
The corresponding single-impurity Anderson model~\cite{Andersonimp} 
and its periodic generalization (Varma--Yafet model~\cite{Varma})
apply, e.g., for the rare-earth materials. 
Since the Coulomb interaction between the $f$~electrons is very large,
one often encounters the situation where the $N_f$-fold degenerate 
levels are at most singly occupied. 
Hence, the issue of degenerate bands with finite electron density
need not be addressed in these cases.

In the past few years the multi-band Hubbard model 
received new attention.
The Gutzwiller approximation to Gutzwiller-correlated variational
wave functions was generalized in~\cite{Lu}--\cite{Buenemann}, 
the Slave Boson mean-field approach along the lines of 
Kotliar and Ruckenstein~\cite{KR}
was developed and used
in~\cite{Dorin}--\cite{Fresard}, and dynamical mean-field methods were
applied in~\cite{Kajueter,Rozenberg}.
In this paper we extend the method in~\cite{Gebhard1,Gebhard2}
to evaluate general Gutzwiller-correlated wave functions
in the limit of infinite space dimensions without further approximations.
In this way we recover the results of~\cite{Lu}--\cite{Buenemann} and thus
show that these earlier results become exact in infinite dimensions within the 
variational approach. Furthermore, we prove that the Slave Boson mean-field
results are variationally controlled at zero temperature in this limit.
Similar results hold for the one-band Hubbard model; 
see~\cite{Revfour} and~\cite[Chap.~3.4,~3.5]{BUCH} for brief reviews.

Our paper is organized as follows. In Section~\ref{wave_functions} 
we introduce a general class
of Gutzwiller-correlated wave functions which allows for the variational study
of Hubbard models with general spin-orbit structure. 
We set up a diagrammatic perturbation theory to
calculate expectation values for the variational ground-state energy 
and other physical quantities of interest.
In Section~\ref{exact_results} we use 
our formalism to derive exact
analytical expressions for the ground-state energy in the limit of
infinite spatial dimensions. In Section \ref{compareit} we
compare them to the results
of approximate treatments of Gutzwiller-correlated wave functions
and those of the Slave Boson mean-field theory.
Since our treatment allows for the inclusion 
of site-diagonal energetic disorder in the Hamiltonian, we
briefly discuss
the consequences of strong correlations on the Anderson transition 
within our variational description.
A summary in Section~\ref{summary} closes our presentation.
\newpage

\section{Gutzwiller-correlated wave functions}
\label{wave_functions}

In this section we introduce the Hubbard Hamiltonian 
for degenerate bands and the class of Gutzwiller-correlated wave functions
as our approximate ground states.
The minimization of the ground-state energy fixes the variational
parameters contained in these trial states.
For the evaluation of the corresponding many-body expectation values 
we develop a diagrammatic formalism
which allows for the complete solution of the problem
in infinite space dimensions.

\subsection{Definitions}

In his first two papers on narrow-band electron systems
Hubbard~\cite{Hubbard,Hubbard2} considered 3$d$~electrons
with a purely local interaction. To simplify our
considerations we make the further assumption that
the interaction depends only on the number densities of electrons
in the $d$~orbitals. Then we may write the Hamiltonian in the form
\begin{equation}
\hat{H}= \sum_{i\neq j;\sigma ,\sigma'}t_{i,j}^{\sigma, \sigma'}
\hat{c}_{i;\sigma }^{+}\hat{c}_{j;\sigma'}^{\phantom{+}}
+\sum_{i;\sigma} \epsilon_{i}^{\sigma} \hat{n}_{i;\sigma }
+ \sum_{i;\sigma,\sigma'} U_i^{\sigma,\sigma'}
\hat{n}_{i;\sigma }\hat{n}_{i;\sigma'} \label{1} \; , 
\end{equation}
where $\sigma,\sigma'=1,\ldots, 2N$ are combined spin-orbit indices
($N=5$ for $d$~electrons),
$i$, $j$ denote lattice sites, and $t_{i,j}^{\sigma,\sigma'}$
is the matrix element for the electron transfer between
two sites~$i$ and~$j$ with spin-orbit indices $\sigma$ and $\sigma'$,
respectively. In the following we will use the notion ``orbital''
for spin-orbital states.

The local energies~$t_{i,i}^{\sigma,\sigma'}=\delta_{\sigma,\sigma'}
\epsilon_{i}^{\sigma}$ can be arbitrarily chosen to describe
different orbital energy levels, e.g., in compounds or 
to mimic the influence of impurities and other lattice defects. 
As we will show below,
expectation values for Gutzwiller-correlated wave functions
can be calculated analytically in infinite dimensions
even in the presence of energetically random impurity potentials.

Finally, $U_i^{\sigma, \sigma'}$ describes the local
Coulomb interaction between two electrons
in the orbitals~$\sigma$ and $\sigma'$
on the same lattice site. The local interaction partly
accounts for the atomic exchange coupling. 
For example, in the case $N=2$ the ground state of an atom with 
two electrons should be a spin triplet according to Hund's first
rule. According to our Hamiltonian~(\ref{1}) the interaction energy 
of the local $S^z=\pm 1$ states is different from 
that of the $S^z=0$ states.
However, the two $S^z=0$ states are still energetically degenerate in~(\ref{1}),
i.e., the interaction does not fully distinguish 
between triplet and singlet states, as required by
Hund's first rule. 
In this work we restrict ourselves to density-dependent
terms only. This should be the generic case for 
C$_{60}$ and other fullerenes for which exchange coupling 
is less important because of the large diameter of the molecules;
see, e.g.,~\cite{Gunnarsson}. 
For other materials it might be necessary to include local spin-flip terms
both in the Hamiltonian and in the variational description.
Since the evaluation of many-body wave functions is
fairly complicated we restrict
ourselves to Gutzwiller correlators which are
solely dependent on the density operators [see~(\ref{5}) below],
and also omit the spin-flip terms in the Hamiltonian for consistency.

We introduce the following notations for the $2^{2N}$ possible configurations 
of a given lattice site.
\begin{enumerate}
\item An atomic configuration~$I$ is characterized by the electron occupation
of the orbitals,
\begin{equation}
\fl I\in  \left\{\emptyset;
(1),\ldots,(2N);
(1,2),\ldots,(2,3),\ldots (2N-1,2N);
\ldots;
(1,\ldots,2N)
\right\}
\;,  \label{2}
\end{equation}
where the order of numbers in $(abc\ldots)$ is irrelevant. The symbol
$\emptyset$ in~(\ref{2}) means that the site is empty.
In general, we interpret the indices~$I$ in~(\ref{2}) as sets in the usual sense.
For example, in the atomic configuration
$I\backslash I'$ 
only those orbitals in~$I$ are occupied which are not in~$I'$.
The complement of~$I$ is 
$\overline{I}=(1,2,\ldots,2N)\backslash I$,
i.e., in the atomic configuration $\overline{I}$ all orbitals but those
in~$I$ are occupied. 
\item The absolute value $|I|$ of a configuration 
is the number of elements in it, i.e.,
\begin{equation}
|\emptyset|=0;|(a)|=1;|(a,b)|=2;\ldots;|(1,\ldots,2N)|=2N
\;.  \label{3}
\end{equation}
\item The operator which projects onto a specific configuration~$I$
on site~$i$ is given by
\begin{mathletters}
\label{g1}
\begin{equation}
\hat{m}_{i;I} =\prod_{\sigma \in I}\hat{n}_{i;\sigma}
\prod_{\sigma \in \overline{I}}(1-\hat{n}_{i;\sigma })
\quad , \quad m_{i;I} = \langle \hat{m}_{i;I}\rangle \;,  \label{4b}
\end{equation}
where $\langle \ldots \rangle$ denotes the expectation value
in the Gutzwiller-correlated wave function; see below.
The operators~$\hat{m}_{i;I}$ measure the ``net'' occupancy.
Besides these we define the operators for the ``gross'' occupancy as
\begin{equation}
\hat{n}_{i;I} =\prod_{\sigma \in I}\hat{n}_{i;\sigma}
\quad ; \quad n_{i;I}= \langle \hat{n}_{i;I} \rangle 
\quad ; \quad \hat{n}_{i;\emptyset} \equiv 1 \;.  \label{4}
\end{equation}\end{mathletters}%
The gross occupancy operator $\hat{n}_{i;I}$ gives a non-zero result
when applied to~$I'$ only if~$I$ contains electrons
in the same orbitals as~$I'$. 
However, $I$ and $I'$ need not be identical because
$I'$ could contain additional electrons in further orbitals, i.e.,
only $I \subseteq I'$ is required.
\end{enumerate}
Each gross (net) operator can be written as a sum of (net) gross operators 
\begin{mathletters}
\label{g2}
\begin{eqnarray}
\hat{n}_{i;I} = \sum_{I'\supseteq I}\hat{m}_{i;I'}
\;,  \label{4c} \\
\hat{m}_{i;I} =\sum_{I'\supseteq I}
\left(-1\right)^{|I'\backslash I|}\hat{n}_{i;I'}
\;.
\label{4d}
\end{eqnarray}\end{mathletters}%
For practical calculations the net 
operators~$\hat{m}_{i;I}$ are more useful than the
gross operators~$\hat{n}_{i;I}$
because the former are projection operators 
onto a given configuration~$I$ on site~$i$, i.e.,
$\hat{m}_{i;I}\, \hat{m}_{i;I'} = \delta_{I,I'} \hat{m}_{i;I}$.

With these definitions we may rewrite
the interaction part of the Hamiltonian~(\ref{1}) as 
\begin{mathletters}
\label{interact}
\begin{equation}
\sum_{i;\sigma ,\sigma'} U_i^{\sigma, \sigma'}
\hat{n}_{i;\sigma }\hat{n}_{i;\sigma'}=
\sum_{i;I\, (|I|\geq 2) } U_{i;I}\, \hat{m}_{i;I}\;,  \label{4e}
\end{equation}
with 
\begin{equation}
U_{i;I}=\sum_{\sigma ,\sigma' \in I} U_i^{\sigma, \sigma'}\;.
\end{equation}\end{mathletters}%
With the help of these definitions 
we may formulate the class of Gutzwiller-correlated
wave functions for degenerate bands as
\begin{mathletters}
\label{5}
\begin{eqnarray}
\left| \Psi_{\rm G}  \right\rangle
=\hat{P}_{\rm G} \left| \Psi_0\right\rangle \label{5a} \; , \\[9pt]
\hat{P}_{\rm G} = \prod_i\prod_{{I}\atop {(|I|\geq 2)}}
g_{i;I}^{\hat{m}_{i;I}} \;.  \label{5b}
\end{eqnarray}\end{mathletters}%
The trial states depend on $K_N=2^{2N}-(2N+1)$ real numbers $g_{i;I}$ 
for each lattice site and on further variational parameters
in $\left| \Psi _0\right\rangle$. 
In general, the gross occupation 
densities $n_{i;\sigma}$ in $|\Psi_{\rm G} \rangle$ 
are different from those in $|\Psi_0\rangle$.
The generalized
Gutzwiller correlator~$\hat{P}_{\rm G}$ in~(\ref{5}) suppresses fluctuations
in the multiple orbital occupancy for repulsive interactions. 
In a translationally invariant system the Gutzwiller wave functions
for degenerate bands as proposed in~\cite{Buenemann} 
are recovered by setting $g_{i;I}\equiv g_{I}$.
Similar but different expressions 
for the Gutzwiller wave function for degenerate
bands can be found in the work by Gutzwiller~\cite{Gutzwiller2},
Gutzwiller and Chao~\cite{ChaoGutz}, and Chao~\cite{Chao}. 

\subsection{Diagrammatic evaluation}

To gain further insight into the physics of the variational
wave functions we have to evaluate expectation values
\begin{equation}
\langle \hat{O} \rangle = 
\frac{ \langle \Psi_{\rm G} | \hat{O}| \Psi_{\rm G} \rangle}%
{ \langle \Psi_{\rm G} | \Psi_{\rm G} \rangle} \; .
\end{equation}
The variational parameters in $|\Psi_{\rm G} \rangle$ are obtained
by the minimization of
the expectation value of the Hamiltonian~(\ref{1}),
\begin{mathletters}
\label{7}
\begin{eqnarray}
E_0^{\rm var} 
= \mathop{\rm min}_{g_{i;I}; |\Psi_0\rangle} \langle \hat{H}\rangle
\; , \label{7a}\\[6pt]
\langle \hat{H}\rangle =
\sum_{i,j;\sigma,\sigma'}t_{i,j}^{\sigma,\sigma'}
\langle\hat{c}_{i;\sigma}^{+}\hat{c}_{j;\sigma'}^{\phantom{+}}\rangle
+\sum_{i;I\,(|I|\geq 2)} U_{i;I}\, \langle \hat{m}_{i;I}\rangle \;.  \label{7b}
\end{eqnarray}\end{mathletters}%
The variational ground-state energy $E_0^{\rm var}$ is an
upper bound for the exact ground-state energy.
This upper-bound property applies only if we are able to
evaluate the variational ground-state energy without
further approximations.

The evaluation of expectation values with correlated wave functions
is a many-particle problem that cannot be solved in general;
see~\cite{Revfour} and~\cite[Chap.~3.4]{BUCH} for a review.
In this paper we will use the method introduced by Gebhard~\cite{Gebhard1,Gebhard2},
which allows the approximation-free evaluation of general Gutzwiller-correlated
wave functions in the limit of infinite spatial dimensions.
The formalism is based on a diagrammatic expansion
of expectation values in such a way that the (variational) self-energy
identically vanishes in infinite dimensions.

As shown in more detail in~\cite{Gebhard1} we have to carry out 
the following program: (i)~choose the appropriate expansion parameters,
(ii)~apply Wick's theorem, and (iii)~use the linked-cluster theorem. 
If the expansion parameters are properly
chosen the lowest order in the expansion already gives the exact result
in infinite dimensions.

\subsubsection{Choice of the expansion parameters.}

As a first step we choose the appropriate expansion parameter(s)
for our diagrammatic theory. In this respect the
variational approach is more flexible than the standard perturbation
theory for Green functions in interacting electron systems.

We express the one-particle product wave function
$|\Psi_0\rangle$ in the form
\begin{equation}
\left| \Psi_0\right\rangle =\prod_i
\eta_{i;\emptyset}^{}\eta_{i;1}^{\hat{n}_{i;1}}\ldots\eta _{i;2N}^{\hat{n}_{i;2N}}
\left| \Phi_0\right\rangle \;,
\label{8}
\end{equation}
where $|\Phi_0\rangle$ is another, normalized
one-particle product wave function.
The real numbers $\eta _{i;\emptyset},\ldots,\eta _{i;2N}$ 
are chosen such
that the square of the ``modified'' (local) Gutzwiller correlator, 
\begin{eqnarray}
\widehat{P}_i \equiv \eta_{i;\emptyset}^{}\prod_{\sigma=1}^{2N}
\eta_{i;\sigma}^{\hat{n}_{i;\sigma}}
\prod_{{I}\atop {(|I|\geq 2)}} g_{i;I}^{\hat{m}_{i;I}}  \nonumber \\[6pt]
=\eta_{i;\emptyset} \prod_{\sigma =1}^{2N} 
\left[ 1+(\eta_{i;\sigma }-1)\hat{n}_{i;\sigma }\right] 
\prod_{{I}\atop {(|I|\geq 2)}}
\left[1+(g_{i;I}-1)\hat{m}_{i;I}\right] \; ,  \label{9}
\end{eqnarray}
can be written in the form 
\begin{equation}
\widehat{P}_i^2=1+x_{i;1,2}\hat{n}_{i;1,2}^{\rm HF}+x_{i;1,3}\hat{n}_{i;1,3}^{\rm HF}
+\ldots+x_{i;1,\ldots,2N}\hat{n}_{i;1,\ldots,2N}^{\rm HF}\;.  \label{10}
\end{equation}
Here, the parameters $x_{i;I}$ are real numbers 
and the Hartree--Fock (HF) operators 
$\hat{n}_{i;I}^{\rm HF}$ are defined as
\begin{equation}
\hat{n}_{i;I}^{\rm HF}=\prod_{\sigma \in I}(\hat{n}_{i;\sigma}-n_{i;\sigma }^0)
  \label{11}  \quad \hbox{for} \quad |I|\geq 1 \; , 
\end{equation}
and $\hat{n}_{i;\emptyset}^{\rm HF}\equiv 1$. Note that
the definition of the Gutzwiller correlator and the Hartree--Fock operators
differs from 
the one given in~\cite{Gebhard1,Gebhard2}. In~(\ref{11}) we introduced
\begin{equation}
n_{i;\sigma }^0=\langle \Phi_0 | \hat{n}_{i;\sigma }| \Phi_0\rangle
\equiv \langle \hat{n}_{i;\sigma}\rangle_0
\label{12}
\end{equation}
as the local densities in orbital~$\sigma$
in the {\em new\/} single-particle product state~$|\Phi_0\rangle$.
Equation~(\ref{10}) poses $2^{2N}$ conditions for the 
$2^{2N}-(2N+1)$ parameters $x_{i;I}$ ($|I|\geq 2$) and the $2N+1$ parameters 
$\eta_{i;\emptyset},\ldots, \eta_{i;2N}$. We will solve for them in
terms of the original variational parameters~$g_{i;I}$ ($|I|\geq 2$) in
the next section.

The parameters $x_{i; I}$ ($|I|\geq 2$) go to zero
for small interaction strengths
since $g_{i; I}(U_i^{\sigma,\sigma'}\to 0) \to 1$.
Hence, we may use them as the 
expansion parameters for a perturbative approach. 
In its diagrammatic formulation the $x_{i; I}$ play the role of
(internal) vertices at which (at least) four lines intersect.
The crucial point in the expansion~(\ref{10}) is the fact that 
there will be no Hartree (``bubble'') diagrams which are of order
unity in all dimensions. Since we are interested in simple expressions
in infinite dimensions, we included their contribution in the expansion
parameters~$x_{i;I}$.
Note that we will make the necessary assumption that local ``Fock terms''
do not occur, i.e., we demand
\begin{equation}
\langle \Phi_0 | \hat{c}_{i;\sigma}^{+} 
\hat{c}_{i;\sigma'}^{\phantom{+}} | \Phi_0 \rangle = \delta_{\sigma,\sigma'} n_{i;\sigma}^0
\label{noFOCK}
\end{equation}
for the one-particle product wave wave function~$|\Phi_0\rangle$.

\subsubsection{Application of Wick's theorem.}

As our second step we formally expand the
expectation values that we need for the
calculation of the variational ground-state energy~(\ref{7}).
With the help of~(\ref{10}) we may write
\begin{mathletters}
\label{g3}
\begin{eqnarray} 
\langle \Psi_{\rm G} | \hat{m}_{f;I} |\Psi_{\rm G}\rangle
=\langle \Phi_0 | \widehat{P}_{f} \hat{m}_{f;I} \widehat{P}_{f}
\prod_{i\neq f} \widehat{P}_i^2 | \Phi_0 \rangle \; ,  \label{13a} \\[6pt]
\langle \Psi_{\rm G} 
| \hat{c}_{f;\sigma_f}^{+}\hat{c}_{h;\sigma_h}^{\phantom{+}}
|\Psi_{\rm G}\rangle =
\langle \Phi _0 | 
\left( \widehat{P}_{f} \hat{c}_{f;\sigma_f}^{+}\widehat{P}_{f}\right) 
\left( \widehat{P}_{h} \hat{c}_{h;\sigma_h}^{\phantom{+}} \widehat{P}_{h}\right) 
\prod_{i\neq f\, (f,h)}\widehat{P}_i^2 | \Phi_0\rangle \; , \label{13c}\\[6pt]
\prod_{i\neq f,h}\widehat{P}_i^2 = 
1+\sum_{k=1}^\infty \frac 1{k!}
\mathop{{\sum}'}_{i_1,\ldots,i_k} \sum_{{I_{i_1},\ldots,I_{i_k}}\atop {(|I_i|\geq 2)}}
\prod_{j=i_1}^{i_k} \left( x_{j; I_j} \hat{n}_{j;I_j}^{\rm HF} \right)
\;,  \label{13b}
\end{eqnarray}
where the primes on a sum indicate that all lattice sites
are different, 
\begin{equation}
f\neq h\neq i_1\neq ...\neq i_k\;.  \label{13e} 
\end{equation}\end{mathletters}%
There are still some Hartree contributions contained 
in~(\ref{13a}) and~(\ref{13c}) which come from the ``external''
sites~$f$ and~$h$.
With the help of~(\ref{g1}) we can always find 
unique expansions of the form 
\begin{mathletters}
\label{g4}
\begin{eqnarray}
\widehat{P}_{f}\hat{m}_{f;I}\widehat{P}_{f}
=\sum_{I_{f}} o_{f;I_{f}}^I \hat{n}_{f;I_{f}}^{\rm HF} \; ,  \label{14}
\\[6pt]
\widehat{P}_{f}\hat{c}_{f;\sigma_f}^{+}\hat{P}_{f} =
\hat{c}_{f;\sigma_f}^{+} \sum_{I_{f} \, (\sigma_f \not\in I_f)}
z_{f;I_{f}}^{\sigma_f} \hat{n}_{f;I_{f}}^{\rm HF}\;,  \label{14b}
\\[6pt]
\widehat{P}_{h}\hat{c}_{h;\sigma_h}^{\phantom{+}}\widehat{P}_{h} =
\hat{c}_{h;\sigma_h}^{\phantom{+}} \sum_{I_{h}\, (\sigma_h \not\in I_h)} 
z_{h;I_{h}}^{\sigma_h}
\hat{n}_{h;I_{h}}^{\rm HF}\;.  \label{14c}
\end{eqnarray}\end{mathletters}%
Later, in Sect.~\ref{energycalc}, we will calculate explicitly
the coefficients~$o_{f;I_f}^I$ and~$z_{f;I_f}^{\sigma_f}$ in terms
of the variational parameters~$g_{i;I}$ and~$\eta_{i;\emptyset}$,
$\eta_{i;\sigma}$.
We introduce the abbreviations
\begin{mathletters}
\label{g6}
\begin{eqnarray}
T_{f,i_1,\ldots,i_k}^{I_{f},I_{i_1},\ldots,I_{i_k}}
=\Bigl\langle \Phi_0\Bigl| 
\hat{n}_{f;I_f}^{\rm HF} \prod_{j=i_1}^{i_k}\hat{n}_{j;I_j}^{\rm HF} 
\Bigr| \Phi_0 \Bigr\rangle \; ,  \label{16} \\[6pt]
S_{f,h,i_1,\ldots,i_k}^{I_{f},I_{h},I_{i_1},\ldots,I_{i_k}} (\sigma_f,\sigma_h)
=\Bigl\langle \Phi_0\Bigl| \hat{c}_{f;\sigma_f}^{+}\hat{c}_{h;\sigma_h}^{\phantom{+}}
\hat{n}_{f;I_f}^{\rm HF}\hat{n}_{h;I_h}^{\rm HF}\prod_{j=i_1}^{i_k}\hat{n}_{j;I_j}^{\rm HF}
\Bigr| \Phi_0\Bigr\rangle  \; ,  \label{17}
\end{eqnarray}\end{mathletters}%
where the products are replaced by unity for $k=0$. This notation
allows us to rewrite~(\ref{g3}) as
\begin{mathletters}
\label{g5}
\begin{eqnarray}
\fl 
\langle \Psi_{\rm G} | \hat{m}_{f;I}| \Psi_{\rm G} \rangle
=\sum_{I_{f}} o_{f;I_{f}}^I 
\Biggl[ T_f^{I_f} + \sum_{k=1}^{\infty} \frac{1}{k!} 
\mathop{{\sum}'}_{i_1,\ldots,i_k}
\sum_{{I_{i_1},\ldots,I_{i_k}}\atop{(|I_i|\geq 2)}}
\biggl( \prod_{j=i_1}^{i_k} x_{j;I_j} \biggr)
T_{f,i_1,\ldots,i_k}^{I_{f},I_{i_1},\ldots,I_{i_k}}  \Biggr]
\;,  \label{15} \\[6pt]
\fl \langle \Psi_{\rm G} |\hat{c}_{f;\sigma_f}^{+}\hat{c}_{h;\sigma_h}^{\phantom{+}}
|\Psi_{\rm G} \rangle 
= \sum_{I_{f},I_{h}} z_{f;I_{f}}^{\sigma_f} z_{h;I_{h}}^{\sigma_h} \nonumber \\
\times \biggl[ S_{f,h}^{I_f,I_h} (\sigma_f,\sigma_h) + \sum_{k=1}^{\infty} \frac{1}{k!}
\mathop{{\sum}'}_{i_1,\ldots,i_k}
\sum_{{I_{i_1},\ldots,I_{i_k}}\atop{(|I_i|\geq 2)}}
\Bigl( \prod_{j=i_1}^{i_k} x_{j;I_j} \Bigr) 
S_{f,h,i_1,\ldots,i_k}^{I_{f},I_{h},I_{i_1},\ldots,I_{i_k}} (\sigma_f,\sigma_h)\biggr] \;.
\nonumber \\[3pt]
\end{eqnarray}\end{mathletters}%
The primes on the lattice sums indicate that all lattice indices 
are different from each other. Moreover, the definition~(\ref{3}) implies that
a given orbital index occurs only once at each lattice site.

Now we are in the position to apply Wick's theorem~\cite{Fetter}. 
The expectation values~(\ref{g6}) can then be written as 
determinants.
Equation~(\ref{16}) becomes 
\begin{equation}
T_{f,i_1,\ldots,i_k}^{I_{f},I_{i_1},\ldots,I_{i_k}}
=\left| 
\begin{array}{cccc}
M_{f,f} & M_{f,i_1} & \cdots & M_{f,i_k} \\ 
M_{i_1,f} & M_{i_1,i_1} & \cdots & M_{i_1,i_k} \\ 
\vdots & \vdots &  & \vdots \\ 
M_{i_k,f} & M_{i_k,i_1} & \cdots & M_{i_k,i_k}
\end{array}
\right| \; ,  \label{18}
\end{equation}
where we introduced the sub-matrices 
\begin{equation}
M_{i,j}=\left( 
\begin{array}{ccc}
P_{i,j}^{\sigma_1,\sigma'_1} & \cdots & P_{i,j}^{\sigma_1,\sigma'_{|I_j|} } \\ 
\vdots &  & \vdots \\ 
P_{i,j}^{\sigma_{|I_i|},\sigma'_1} & \cdots & P_{i,j}^{\sigma_{|I_i|},\sigma'_{|I_j|}}
\end{array}
\right) 
\;.  \label{19}
\end{equation}
Here, the orbital indices $\sigma_1,\ldots,\sigma_{|I_i|}$ 
($\sigma'_1,\ldots,\sigma'_{|I_j|}$)
are the elements of $I_i$ ($I_{j}$) and 
\begin{equation}
P_{i,j}^{\sigma,\sigma'}=(1-\delta_{i,j})
\langle \Phi_0 | \hat{c}_{i;\sigma}^{+}\hat{c}_{j;\sigma'}^{\phantom{+}} 
| \Phi_0\rangle 
\equiv (1-\delta_{i,j}) 
\langle \hat{c}_{i;\sigma}^{+}\hat{c}_{j;\sigma'}^{\phantom{+}} \rangle_0 
 \label{20}
\end{equation}
is the one-particle density matrix for the
one-particle product wave function $|\Phi_0\rangle$ for
$i\neq j$. These objects play the role of ``lines''
in our diagrammatic expansion. Note that we do not have to distinguish
between ``hole'' and ``particle'' lines because all sites
are different when we apply Wick's theorem~\cite{Gebhard1}--\cite{Revfour}.
Furthermore, local terms did not
arise because we subtracted the Hartree contributions 
and ruled out Fock terms according to~(\ref{noFOCK}).
Similarly, equation~(\ref{17}) becomes
\begin{equation}
\fl S_{f,h,i_1,\ldots,i_k}^{I_{f},I_{h},I_{i_1},\ldots,I_{i_k}}(\sigma_f,\sigma_h)
=(-1)^{|I_f| |I_h|-|I_f|-|I_h|}
\left| 
\begin{array}{ccccc}
M_{h,f} & M_{h,h} & M_{h,i_1} & \cdots & M_{h,i_k}
\\ 
M_{f,f} & M_{f,h} & M_{f,i_1} & \cdots & M_{h,i_k}
\\ 
M_{i_1,f} & M_{i_1,h} & M_{i_1,i_1} & \cdots & M_{i_1,i_k}
\\ 
\vdots & \vdots  & \vdots  &  & \vdots  \\ 
M_{i_k,f} & M_{i_k,h} & M_{i_k,i_1} & \cdots & M_{i_k,i_k}
\end{array}
\right| \;.  \label{25}
\end{equation}
The matrices $M_{i,j}$ in~(\ref{25}) are again given by~(\ref{19}). Here,
the orbital indices belonging to the lattice sites~$f$ and~$h$ are elements
of $I_f$, $I_f\cup \sigma_f$, $I_h$, and $I_h \cup \sigma_h$, respectively.

To calculate expectation values for~$|\Psi_{\rm G} \rangle$
we have to divide the numerators, (\ref{g5}), by the norm 
\begin{equation}
\langle \Psi_{\rm G} | \Psi_{\rm G} \rangle =\langle \Phi_0|\Phi_0\rangle 
+ \sum_{k=1}^\infty \frac{1}{k!}
\mathop{{\sum}'}_{i_1,\ldots,i_k}
\sum_{{I_{i_1},\ldots,I_{i_k}}\atop{(|I_i|\geq 2)}}
\Bigl( \prod_{j=i_1}^{i_k} x_{j;I_{j}}\Bigr)
T_{i_1,\ldots,i_k}^{I_{i_1},\ldots,I_{i_k}} \; .  \label{26}
\end{equation}
In principle, we could derive diagram rules for the series expansion
of the determinants 
in powers of the parameters~$x_{j;I_j}$. In this paper we restrict ourselves
to the limit of infinite dimensions where not a single diagram needs
to be calculated.

\subsubsection{Linked-cluster theorem.}

The summation restrictions prevent us from
the application of the linked-cluster theorem.
In the case of a single band ($N=1$)
the summation restrictions~(\ref{13e}) can simply be dropped
because the determinants~(\ref{18}) and~(\ref{25}) vanish identically 
if two lattice indices coincide. Since we may then independently sum
over all lattice sites the linked-cluster theorem applies~\cite{Fetter}
such that the disconnected diagrams in the numerator are canceled by
the norm~\cite{Gebhard1}.

The case $N\geq 2$ requires more care because, in general, 
the determinants $T$ in~(\ref{18}) and $S$ in~(\ref{25}) remain
finite when we equate two lattice sites.
To make progress we note that a
summation restriction over the spin-orbit indices in
$I_{i_1},\ldots,I_{i_k}$ will not prevent
the applicability of the linked-cluster theorem. To see this, we write
a typical spin-orbit sum in~(\ref{g5}) in the form~\cite{Gebhard2}
\begin{equation}
\sum_I f(I)=\sum_{r=1}^{2N} \frac{1}{r!}
\mathop{{\sum}'}_{\alpha _1,\ldots,\alpha_r} f(\alpha_1,\ldots,\alpha_r)\;.  \label{27}
\end{equation}
The ensuing summation restriction 
$\alpha _1\neq \alpha _2\neq \ldots\neq \alpha_r$ 
can be removed because the determinants in~(\ref{18}) and~(\ref{25}) vanish
if two of these indices are identified. 

Now we equate two lattice sites~$l$ and~$m$, e.g., in the
determinant~(\ref{18}). 
Two different cases have to be distinguished: 
(i)~if $I_l$ and $I_m$ have at least two 
elements in common the determinant~(\ref{18}) 
vanishes since at least two rows or columns will be identical;
(ii)~otherwise, if $I_l$ and $I_m$ have no common element, the
determinant~(\ref{18}) remains finite. 
When we equate two lattice sites in the determinant we effectively map two
vertices onto each other in the corresponding diagrammatic expansion,
i.e., we generate a diagram that already appeared at some lower order
in the expansion; see~\cite{Gebhard2} for a simple example for this
``vertex packing''.
The original diagram with $l$ and $m$ put equal and the corresponding
lower-order diagram have the same topology but different prefactors,
$x_{l;I_l\cup I_m}$ for the lower-order diagram and 
$x_{l;I_l}x_{m;I_m}$ for the new contribution.
If we relax all summation restrictions in the numerator and in
the norm we must replace
the vertices $x_{i;I_i}$ and the coefficients $o_{i;I_i}^I$ 
and $z_{i;I}^{\sigma}$ by effective ones, i.e.,
$x_{i;I}\rightarrow \widetilde{x}_{i;I}$, $o_{i;I_i}^I \rightarrow 
\widetilde{o}_{i;I_i}^I$, and $z_{i;I}^{\sigma} \rightarrow 
\widetilde{z}_{i;I}^{\sigma}$. 
Fortunately, the lowest-order coefficients $o_{i;\emptyset}^I$ and 
$z_{i;\emptyset}^{\sigma}$ remain unchanged by this procedure
because these coefficients correspond to diagrams without an internal
vertex whereas the ``vertex packing'' leaves behind at least one internal 
vertex~\cite{Gebhard2}.
In the next section we will show 
that~$o_{i;\emptyset}^I$ and~$z_{i;\emptyset}^{\sigma_i}$ alone give the
exact result in infinite dimensions.
Since the number of correction terms generated in a given 
order of the expansion remains finite, 
the calculation of systematic $1/d$ corrections is still possible 
although very tedious.

Now that we formally eliminated all summation restrictions
we can apply the linked-cluster theorem.
We thus find
\begin{mathletters}
\label{g7}
\begin{eqnarray}
\fl \langle \hat{m}_{f;I}\rangle
=o_{f;\emptyset}^I +\sum_{{I_{f}}\atop {(|I_f|\geq 1)}}
\widetilde{o}_{f;I_{f}}^I
\sum_{k=1}^\infty \frac{1}{k!} \sum_{i_1,\ldots,i_k}
\sum_{{I_{i_1},\ldots,I_{i_k}}\atop{(|I_i|\geq 2)}} 
\biggl\{ \hat{n}_{f;I_f} \prod_{j=i_1}^{i_k}
\widetilde{x}_{j;I_{j}}
\hat{n}_{j;I_{j}}^{\rm HF}\biggr\}_0^{\rm C}
\label{28} \\[9pt]
\fl \langle \hat{c}_{f;\sigma_f}^{+}\hat{c}_{h;\sigma_h}^{\phantom{+}}
\rangle =
z_{f;\emptyset}^{\sigma_f}z_{h;\emptyset}^{\sigma_h}
\langle \hat{c}_{f;\sigma_f}^{+}\hat{c}_{h;\sigma_h}^{\phantom{+}}\rangle_0
+\sum_{{I_{f},I_{h}}\atop{(|I_{f}|+|I_{h}|\geq 1)}}
\widetilde{z}_{f;I_{f}}^{\sigma_f}\widetilde{z}_{h;I_{h}}^{\sigma_h}
\nonumber \\[6pt]
\lo \times \sum_{k=1}^\infty \frac{1}{k!}\sum_{i_1,\ldots,i_k}
\sum_{{I_{i_1},\ldots,I_{i_k}}\atop {(|I_i|\geq 2)}}
\biggl\{ \hat{c}_{f;\sigma_f}^{+}\hat{c}_{h;\sigma_h}^{\phantom{+}}
\hat{n}_{f;I_{f}}^{\rm HF}\hat{n}_{h;I_{h}}^{\rm HF}
\prod_{j=i_1}^{i_k} \widetilde{x}_{j;I_{j}}\hat{n}_{j;I_{j}}^{\rm HF}
\biggr\}_0^{\rm C}
\;,  \label{29} 
\end{eqnarray}\end{mathletters}%
where, as usual, 
$\left\{ \ldots\right\}_0^{\rm C}$ indicates that only the connected
diagrams need to be considered.

\section{Exact results in infinite dimensions}
\label{exact_results}

In this section we briefly review
the diagrammatic simplifications which occur in infinite dimensions.
We find that in our approach not a single diagram needs to be calculated
in this limit.
Consequently, we derive explicit analytic expressions
for the variational ground-state energy, the one-particle
density matrices, and the average net occupation densities
which are exact in infinite dimensions
and valid for the whole class of Gutzwiller-correlated 
wave functions.

\subsection{Simplifications}

Systematic studies of the limit of infinite dimensions for
itinerant electron systems started with the work of Metzner and
Vollhardt~\cite{MVdinfty}; for details and a recent review, 
see~\cite[Chap.~5]{BUCH}. 
One of the essential simplifications is the following:
if two vertices~$l$ and~$m$ are connected by
three independent (Green function) 
lines only the contribution for $l=m$ survives
in infinite dimensions. For example, 
in the one-band case the (proper) self-energy
becomes purely local in this limit. 
In our variational theory
the lines between two sites represent
the one-particle density matrices $P_{l,m}^{\sigma,\sigma'}$~(\ref{20}).
They {\em vanish\/} by construction, if we set $l=m$.
Consequently, the variational (proper) self-energy is 
identically zero in the limit of infinite dimensions in the 
one-band case~\cite{Gebhard1,Gebhard2}.
Note that we guaranteed in our expansion that there 
are no Hartree contributions
such that all (internal) vertices are connected by at least three
lines. For the case $N\geq 2$ vertices with
more than four lines appear for which our arguments particularly
apply, and any diagram with more than one line must vanish in infinite dimensions.

Consequently, not a single diagram needs to be calculated. Instead, we immediately
find from~(\ref{g7})
\begin{mathletters}
\label{g8}
\begin{eqnarray}
\langle \hat{m}_{f;I}\rangle^{(d=\infty)} =
o_{f;\emptyset}^I   \; , 
\label{31}
\\[6pt]
\langle \hat{c}_{f;\sigma_f}^{+}\hat{c}_{h;\sigma_h}^{\phantom{+}}
\rangle^{(d=\infty)}
= z_{f;\emptyset}^{\sigma_f} z_{h;\emptyset}^{\sigma_h}
\langle \hat{c}_{f;\sigma_f}^{+}\hat{c}_{h;\sigma_h}^{\phantom{+}}\rangle_0  
\label{31b} 
\end{eqnarray}\end{mathletters}%
in infinite dimensions. 
Thus, the problem is formally solved since only the coefficients 
$o_{f;\emptyset}^I$ and $z_{f;\emptyset}^{\sigma_f}$ and
the properties of the single-particle product wave function~$|\Phi_0\rangle$
enter the final expressions. 

\subsection{Explicit results for the ground-state energy}
\label{energycalc}

Thus far the variational ground-state energy for
general Gutzwiller-correlated wave functions 
could only be accomplished for $N=1$~\cite{Gebhard1,Revfour}.
Here we show that explicit expressions for the
ground-state energy can be obtained for all interaction strengths
and all $N\geq 1$.
In the following considerations we suppress the spatial index.

The remaining problem is the calculation of the coefficients 
$o_{i;\emptyset}^I$ and $z_{i;\emptyset}^{\sigma}$. 
We recast the modified Gutzwiller-correlator~(\ref{9}) into the form
\begin{equation}
\widehat{P}=\eta_{\emptyset} 
\left[ 1+\sum_I\left( \widetilde{\eta}_I g_I-1\right) 
\hat{m}_I\right] \;,  \label{33}
\end{equation}
where we introduced the notation
\begin{equation}
\begin{array}{@{}lcll@{}}
\widetilde{\eta}_{\emptyset} =1 \; , \\[6pt]
\widetilde{\eta}_I = \prod\limits_{\sigma \in I} \eta_{\sigma} & \mbox{for} & 
|I|\geq 1 &, \\[6pt]
g_I =1 & \mbox{for} & |I|\leq 1 & .
\end{array}
\end{equation}
Then, the left-hand side of~(\ref{14}) becomes
\begin{equation}
\widehat{P} \hat{m}_I \widehat{P}=\eta_{\emptyset}^2 \widetilde{\eta}_I^2
g_I^2\hat{m}_I \; . 
\end{equation}
Now we expand the net occupancy operators $\hat{m}_I$ in terms of
the Hartree--Fock operators $\hat{n}_{I}^{\rm HF}$ as 
\begin{eqnarray}
\hat{m}_I =\prod_{\sigma \in I}
\left( n_{\sigma}^0 + \hat{n}_{\sigma}^{\rm HF} \right) 
\prod_{\sigma \in \overline{I}}
\left( (1-n_{\sigma}^0) -\hat{n}_{\sigma}^{\rm HF}\right) \nonumber \\[6pt]
 = \sum_{I'} \biggl[ (-1)^{|\overline{I}\cap I'|}
\prod_{\sigma \in I\backslash I'} n_{\sigma}^0
\prod_{\sigma \in \overline{I}\backslash I'} 
\left(1-n_{\sigma}^0\right) \biggr] 
\hat{n}_{I'}^{\rm HF}\;.  \label{35}
\end{eqnarray}
We compare this expression with the right-hand side of~(\ref{14})
and find the simple result
\begin{equation}
o_{I'}^I=(-1)^{|\overline{I}\cap I'|} \eta_{\emptyset}^2 \widetilde{\eta}_I^2 g_I^2 
\prod_{\sigma \in I\backslash I'} n_{\sigma}^0 
\prod_{\sigma \in \overline{I}\backslash I'} \left(1-n_{\sigma}^0\right) \;.  
\label{36}
\end{equation}
The representation of the modified Gutzwiller correlator in~(\ref{33}) allows us to
write
\begin{equation}
\widehat{P}^2=\eta_{\emptyset}^2\biggl[ 1+\sum_I\left( \widetilde{\eta}_I^2 g_I^2-1\right) 
\hat{m}_I \biggr] \; .   \label{42}
\end{equation}
Again, we expand the net occupancy operators $\hat{m}_I$ 
in terms of the Hartree--Fock operators $\hat{n}_{I}^{\rm HF}$~(\ref{35}).

Now we are in the position to compare~(\ref{42}) with~(\ref{10}). 
The constant coefficient gives
\begin{mathletters}
\label{zerothorder}
\begin{equation}
\eta_{\emptyset}^2\left[1 +\sum_{I}
\left(\widetilde{\eta}_I^2 g_I^2-1\right) m_I^0 \right] =1\;,
\end{equation}
where
\begin{equation}
m_I^0= \prod_{\sigma \in I} n_{\sigma}^0
\prod_{\sigma \in \overline{I}} \left( 1-n_{\sigma}^0\right) 
\label{mproddef}
\end{equation}\end{mathletters}%
is the corresponding net occupation density
in the uncorrelated wave function $|\Phi_0\rangle$.
By construction, the coefficient to first order
in~$\hat{n}_{\sigma}^{\rm HF}$ is zero in~(\ref{10}).
We thus find 
\begin{equation}
\eta_{\emptyset}^2\sum_{I\, (\sigma\in I)}
\biggl( \widetilde{\eta}_I^2 g_I^2-1\biggr) 
\frac{m_I^0}{n_{\sigma}^0}
-\eta_{\emptyset}^2 \sum_{I\, (\sigma\not\in I)}
\biggl( \widetilde{\eta}_I^2 g_I^2-1\biggr) 
\frac{m_I^0}{1-n_{\sigma}^0} =0\;.  \label{43}
\end{equation}
Finally, the higher orders in the Hartree--Fock operators
in~(\ref{42}) and~(\ref{10}) fix the coefficients $x_I$, 
\begin{equation}
x_I=\sum_{I^{\prime }}\frac{\widetilde{\eta }_{I^{\prime }}^2\,g_{I^{\prime
}}^2-1}{\widetilde{\eta }_{I^{\prime }}^2\,g_{I^{\prime }}^2}o_I^{I^{\prime
}}\;,  \label{45}
\end{equation}
where we used~(\ref{36}).
For $N\geq 2$ one cannot explicitly solve~(\ref{zerothorder}),
(\ref{43}), and~(\ref{45}) for $\eta_{\emptyset}$, $\eta_{\sigma}$, and~$x_I$
in terms of $n_{\sigma}^0$ and~$g_I$. 

In the limit of infinite dimensions we can obtain explicit results for the
ground-state energy. From~(\ref{31}) we find  
\begin{equation}
m_I =\langle \hat{m}_I\rangle = o_{\emptyset}^I
=\eta_{\emptyset}^2 \widetilde{\eta}_I^2 g_I^2 m_I^0  \label{37}
\label{32}
\end{equation}
in infinite dimensions.
Equation~(\ref{37}) implies that the $\eta$-terms can be expressed
by the net occupancy densities as
\begin{equation}
\eta_{\emptyset}^2 = \frac{m_{\emptyset}}{m_{\emptyset}^0} 
\quad , \quad 
\eta_{\sigma}^2 = 
\frac{m_{\sigma}}{m_{\sigma}^0} 
\frac{m_{\emptyset}^0}{m_{\emptyset}} 
\;. 
 \label{38}
\end{equation}
Note that these simple results do not hold in terms
of the single-particle product wave function~$|\Psi_0\rangle$ in~(\ref{5a})
but only in terms of~$|\Phi_0\rangle$; compare equation~(\ref{8}).
Furthermore, equation~(\ref{37}) allows us to replace the 
original variational parameters~$g_I$ by
their physical counterparts,
the net occupancy densities~$m_I$ ($|I|\geq 2$), as
\begin{eqnarray}
g_{a,b}^2 =\frac{m_{ab} m_{\emptyset}}{m_a m_b} \; , \nonumber \\[3pt]
\vdots \label{g9}\label{39} \\[3pt]
g_{1,\ldots,2N}^2= 
\frac{\left(m_{\emptyset}\right)^{n-1} m_{1,\ldots,2N}}%
{m_1 \cdots m_{2N}} \; .  \nonumber 
\end{eqnarray}
Equations~(\ref{39}) are well known from the 
Gutzwiller approximation for the one-band~\cite{Gutzi63,Vollrev}
and the multi-band case~\cite{Buenemann}.
They show that the parameters~$g_I^2$ rule the law-of-mass action
between single occupancies of a site on the one hand
and its multiple occupancies and vacancies on the other hand.

In infinite dimensions 
equation~(\ref{zerothorder}) is trivially fulfilled since
\begin{equation}
m_{\emptyset}=1 -\sum_{I\, (|I|\geq 1)} m_I
\end{equation}
is true by definition. From the 
definition of net and gross operators~(\ref{g1})
we know that
the (gross) occupancy in the orbital~$\sigma$ is given by
\begin{mathletters}
\begin{equation}
n_{\sigma} =\sum_{I\, (\sigma \in I)} m_I \;.  \label{40}
\end{equation}
We further use
\begin{equation}
\sum_{{I\, (\sigma \not\in I)}\atop { |I|\geq 1} } m_I 
= 1 - m_{\emptyset} - n_{\sigma}
\end{equation}\end{mathletters}%
in~(\ref{43}). With the help of~(\ref{38}) it is easy to show that
\begin{equation}
n_{\sigma} =n_{\sigma}^0  \label{44}
\end{equation}
holds in infinite dimensions. It is thus seen that the
local densities in the Gutzwiller-correlated 
wave function~$|\Psi_{\rm G} \rangle$
and in the one-particle product wave function~$|\Phi_0\rangle$ are the same.
This had not been so if we had worked with
the one-particle product wave function~$|\Psi_0\rangle$, see~(\ref{5a}).
For this reason we introduced the new one-particle 
state~$|\Phi_0\rangle$ in~(\ref{8}),
where the single-particle $\eta$~terms could be
interpreted as ``chemical potentials'' which guarantee that the
average single-orbital occupancies remain unchanged 
by the modified Gutzwiller correlator~(\ref{9}).

Thus far we replaced the original variational parameters~$g_I$ by
their physical counterparts, the average multiple-occupancies~$m_I$.
As a last step we have to express the coefficients $z_{\emptyset}^{\sigma}$
in terms of our new variational parameters.
According to~(\ref{7b}) and~(\ref{31b}) they can be interpreted
as site-dependent renormalization factors for the electron transfer 
between two sites.
For their derivation we introduce the operators
\begin{equation}
\hat{m}_I^{\sigma}= 
\prod_{\sigma' \in I\backslash\sigma} \hat{n}_{\sigma'} 
\prod_{\sigma' \in \overline{I}\backslash\sigma} (1-\hat{n}_{\sigma'}) 
\; . 
\end{equation}
Then we may write
\begin{eqnarray}
\widehat{P} \hat{c}_{\sigma}^+ \widehat{P} =
\eta_{\emptyset} 
\biggl(1 + \sum_{I\, (\sigma \in I)} \left(\widetilde{\eta}_I g_I-1\right) \hat{m}_I\biggr)
\, \hat{c}_{\sigma}^{+}  \, 
\eta_{\emptyset} 
\biggl(1 + \sum_{I\, (\sigma \not\in I)} \left(\widetilde{\eta}_I g_I-1\right) \hat{m}_I\biggr) 
\nonumber 
\\[6pt]
= \eta_{\emptyset}^2\hat{c}_{\sigma}^{+}
\biggl(1 + \sum_{I\, (\sigma \not \in I)} 
\left(\widetilde{\eta}_{I\cup \sigma} g_{I\cup \sigma}-1\right) 
\hat{m}_I^{\sigma}\biggr)
\biggl(1 + \sum_{I\, (\sigma \not\in I)} 
\left(\widetilde{\eta}_I g_I-1\right) \hat{m}_I^{\sigma}\biggr)
\\[6pt]
= \eta_{\emptyset}^2\hat{c}_{\sigma}^{+} \biggl( 1 +
\sum_{I\, (\sigma\not\in I)}
\left( \widetilde{\eta}_{I\cup\sigma} g_{I\cup \sigma} \widetilde{\eta}_I g_I-1\right) 
 \hat{m}_I^{\sigma}\biggr) \; .\nonumber 
\end{eqnarray}
With the help of~(\ref{35}) we may again expand the projection 
operators~$\hat{m}_I^{\sigma}$ in terms of the Hartree--Fock 
operators~$\hat{n}_{I}^{\rm HF}$.
A comparison with the definition~(\ref{14b}) then gives
\begin{equation}
\fl z_I^{\sigma} =\eta_{\emptyset}^2 \delta_{I,\emptyset}+\eta_{\emptyset}^2
\sum_{I' (\sigma \not\in I')} \left(-1\right)^{|I\cap \overline{I'}|}
\left( \widetilde{\eta}_{I'\cup \sigma}g_{I'\cup \sigma}
\widetilde{\eta}_{I'} g_{I'} -1\right)
\prod_{\sigma'\in I'\backslash I} n_{\sigma'}^0
\prod_{\sigma'\in \overline{I'}\backslash (I\cup \sigma) }
\left( 1-n_{\sigma'}^0\right)   \label{47} \, .
\end{equation}
This includes the special case of $I=\emptyset$ which we need
for the calculation of the renormalization factors for the electron
transfers,
\begin{equation}
z_{\emptyset}^{\sigma} =
\eta_{\emptyset}^2
+\eta_{\emptyset}^2 \sum_{I\, (\sigma \not\in I)}
\left( \widetilde{\eta}_{I\cup \sigma } g_{I\cup \sigma }
\widetilde{\eta}_I g_I-1\right) \frac{m_I^0}{1-n_{\sigma}^0}\; .
\end{equation}
We use~(\ref{37}) and the identities
\begin{mathletters}
\begin{eqnarray}
m_{I\cup\sigma}^0 =  \frac{n_{\sigma}^0}{1-n_{\sigma}^0} m_I^0 \; , 
\\[6pt]
\sum_{I\, (\sigma\not\in I)} m_I^0 = 1-n_{\sigma}^0 \; ,
\end{eqnarray}\end{mathletters}%
see~(\ref{mproddef}) and~(\ref{40}), and finally arrive at 
\begin{equation}
\sqrt{q_{\sigma}} \equiv z_{\emptyset}^0 
=\frac{1}{\sqrt{\left( 1-n_{\sigma}^0\right) n_{\sigma}^0}}
\sum_{I\, (\sigma\not\in I)}
\sqrt{m_{I\cup \sigma} m_I}\;.  \label{48}
\end{equation}
This leads to our final expression for the ground-state energy 
\begin{equation}
E_0^{\rm var}\left(m_{i;I};|\Phi_0\rangle \right)
= \sum_{i,j;\sigma,\sigma'}
\sqrt{q_{i;\sigma}}\sqrt{q_{j;\sigma'}} t_{i,j}^{\sigma,\sigma'}
\langle \hat{c}_{i;\sigma}^{+}\hat{c}_{j;\sigma'}^{\phantom{+}}\rangle_0
+\sum_{i;I\, (|I|\geq 2)} U_{i;I} m_{i;I}\; .  \label{49}
\end{equation}
To fix the variational parameters~$m_{i;I}$ ($|I|\geq 2$)
and those in~$|\Phi_0\rangle$ the minimum of this
expression has to be determined.
Applications of the formalism are planned to be published elsewhere.

\section{Comparison with other approaches}
\label{compareit}

In this section we compare our results to those of related work, namely
the Gutzwiller approximation and the Slave Boson mean-field
approach, which are seen to become variational controlled in
infinite dimensions.
Finally, we briefly discuss the applicability of the variational approach
to the Anderson transition in strongly correlated electron systems.

\subsection{Generalized Gutzwiller Approximations}

For the one-band case it is known that the results of the Gutzwiller
approximation for the Gutzwiller wave function become exact in
infinite dimensions~\cite{Gebhard1,MVdinfty,Metzner}. 
In addition, the variational calculation of
correlation functions is straightforward in infinite 
dimensions~\cite{Gebhard1}, whereas 
their consistent treatment is beyond the Gutzwiller approximation
scheme.

Generalizations of the Gutzwiller approximation
to other situations than the translationally invariant, paramagnetic
one-band Gutzwiller wave function encounter two conceptual difficulties.
(i)~It is not clear from the beginning 
whether the generalized Gutzwiller approximations 
in~\cite{Lu}--\cite{Buenemann} lead to physically acceptable
results in the whole parameter space.
For example,
early extensions of the Gutzwiller approximation
to the case of antiferromagnetism
gave rise to negative
occupation densities~\cite{Ogawa}.
(ii)~The Gutzwiller approximation may give the correct results in infinite dimensions 
but the corresponding wave function could not be identified properly.
For example, in a recent treatment Okabe~\cite{Okabe}
found the correct expression for the ground-state energy~(\ref{49})
in infinite dimensions for the translationally invariant case but
he did not specify the corresponding Gutzwiller wave function.
Therefore, his results could not be tested against numerical approaches. 
In contrast, the wave function~(\ref{5}) provides a solid starting point
for variational Monte-Carlo simulations. Hence, an assessment of
the quality of the Gutzwiller approximation for finite dimensions
can now be performed; see, e.g.,~\cite{Yoko,Michael} for 
applications to the one-band case. 

The proper choice of the wave function~(\ref{5})
was essential for its approximation-free
evaluation in infinite dimensions.
In our definition of
generalized Gutzwiller-correlated wave functions 
{\em each\/} multiple occupancy
of a lattice site is controlled by its own variational parameter. 
Equations~(\ref{39}) 
show that we may then replace these variational parameters by their physical
counterparts, the net multiple occupancies~$m_I$. 
In contrast, the trial state of
Chao and Gutzwiller~\cite{ChaoGutz} and
Chao~\cite{Chao}
only includes variational 
parameters for the gross occupancies $n_I$ for $|I|=2$,
\begin{equation}
\left| \Psi^{\prime }_{{\rm G}}\right\rangle=
\prod_i\prod_{{I}\atop{(|I|=2)}}g_I^{\hat{n}_{i;I}}|\Psi_0\rangle\;,  \label{50}
\end{equation}
where we used our notation.
In their evaluation the ``maximum term conditions'' of the 
Gutzwiller approximation~\cite{Gutzwiller2} 
also lead to a set of $2^{2N}-(2N+1)$ equations similar to~(\ref{39}) but
it cannot be solved explicitly because there are only $N(2N-1)$ 
variational parameters 
$g_I$ ${(|I|=2)}$ in~(\ref{50}). 
To make progress Chao set to zero all terms with multiple occupancies $|I|\geq 3$,
an assumption not warranted by the wave function~(\ref{50}).
If we (artificially) put $m_{i;I} =0$ for $|I|\geq 3$ for our
wave function~(\ref{5}) we recover Chao's results.
Hence, Chao's results do apply to the Gutzwiller-correlated
wave function~(\ref{5})
in a special limit.

Similarly, the results of Lu~\cite{Lu} are found to be correct if his
further assumptions are adopted. He used site and orbital-independent
Coulomb interactions, $U_i^{\sigma,\sigma'}=U$, and assumed that some of
the multiple occupancies vanish identically. In~\cite{Buenemann} it was shown
that the latter assumption might not hold close to the Brinkman--Rice transition
on the metallic side.

The proper form of the wave function
for arbitrary band degeneracies~(\ref{5})
was first given by B\"unemann and Weber~\cite{Buenemann}
for translationally invariant systems~($g_{i;I}\equiv g_I$).
Using the Gutzwiller approximation they derived
the equations~(\ref{39}),~(\ref{48}), and~(\ref{49}).
In this work we showed that their results become exact in infinite
dimensions. Furthermore, we explicitly covered all cases of symmetry breaking
in the one-particle product wave functions~$|\Phi_0\rangle$; the cases of
spin and orbital ordering in materials with degenerate bands
can equally be studied with the help of Gutzwiller-correlated wave functions.

\subsection{Slave Boson Mean-Field Theory}

For the one-band case the Slave Boson mean-field 
theory~\cite{KR},~\cite{Barnes}--\cite{ReadNewns}
yields the same results as 
Gutzwiller-correlated variational wave functions in infinite
dimensions, including the cases of broken symmetry~\cite{Gebhard1,Gebhard2}.
Therefore, the Slave Boson approach is variationally
controlled at zero temperature in this limit.
The Slave Boson saddle-point free energy
can be re-derived from a ``variational partition 
function''~\cite{Gebhard1,Gebhard2}.
This construction shows that, at best, 
the Slave Boson mean-field theory is applicable
up to excitation energies for which Fermi-liquid theory is valid. 
Since Fermi-liquid parameters can be
derived from the variational ground-state energy~\cite{Vollrev},
the low-energy properties
can equally well be described with the variational and the
Slave Boson mean-field approach.

Recently, the Slave Boson mean-field theory for degenerate
Hubbard models was worked out by Hasegawa~\cite{Hasegawa}
who used the extension of the Kotliar--Ruckenstein
approach for the degenerate Anderson model
by Dorin and Schlottmann~\cite{Dorin}.
His results for $N=2$ completely agree with ours and, thus, 
the above statements on the virtues and limitations of the Slave Boson
mean-field approach also apply for the case of degenerate bands.
Independently, Fr\'esard and Kotliar~\cite{Fresard} derived
a set of Slave Boson mean-field equations and, as an application,
reproduced Lu's results on the Mott transition~\cite{Lu}.

\subsection{Anderson Transition in Strongly Correlated Electron Systems}

Our approach also covers the case
of local, energetically random impurity 
potentials~$\epsilon_i^{\sigma}$.
For illustrative purposes we restrict ourselves to $N=1$, $U=\infty$,
and no spin-flip hopping. 
In this case the system is a Mott insulator
at half band-filling, $\delta=0$, where $\delta=1-(1/L)\sum_i n_i$
is the doping degree of the lower Hubbard band~\cite{BUCH}.
In this case the variational ground-state energy becomes
\begin{mathletters}
\label{gslok}
\begin{eqnarray}
E_0^{\rm var} (\Phi_0) = \sum_{i\neq j;\sigma}t_{i,j}^{\sigma,\sigma}
\sqrt{q_{i;\sigma}}\sqrt{q_{j;\sigma}} 
\langle \Phi_0 | \hat{c}_{i;\sigma}^+\hat{c}_{j;\sigma}^{\phantom{+}}
| \Phi_0 \rangle +\sum_{i} \epsilon_i n_i \; , \\
q_{i;\sigma} =\frac{1-n_i}{1-n_{i;\sigma}} \; . 
\end{eqnarray}\end{mathletters}%
Recall that the local occupation densities 
$n_{i}=n_{i;\sigma}+n_{i;-\sigma}\leq 1$
are given by $n_{i;\sigma}=\langle \hat{n}_{i;\sigma}\rangle =\langle \Phi_0
| \hat{n}_{i;\sigma} | \Phi_0\rangle$.

We may {\em interpret\/} our variational results~(\ref{gslok}) in
terms of a two-fluid picture which naturally arises 
for strongly correlated disordered electron systems. 
A recent introduction and overview 
on the theory of the Anderson transition in interacting electron systems
is given in~\cite{Logan}.
(i)~A certain fraction of electrons may localize, e.g.,
a $\sigma$~electron is localized on site~$l$
in $|\Phi_0\rangle$, $n_{l}=n_{l;\sigma}=1$. From 
the expectation value for the ground-state energy~(\ref{gslok})
it follows that
the probability is zero that the electron hops off the site~$l$.
In addition, a $-\sigma$~electron will not hop onto this site because
this is dynamically forbidden, $q_{l;-\sigma}=0$. Within our variational
description the occupation of site~$l$ does not change.
The site~$l$ is equally likely occupied by a $\sigma$ or a $-\sigma$~electron
and, therefore, the site contributes to a Curie-like susceptibility.
(ii)~The excluded sites represent an unretarded (random) 
hard-core potential for the remaining
electrons besides the fluctuating 
local potentials~$\epsilon_i$ and the dynamical
constraint of no double occupancy. 
The strength of the impurity potentials~$\epsilon_i$ and the degree
of doping determine the fraction of ``localized'' and ``mobile'' electrons.

For a small doping of the lower 
Hubbard band, $\delta \ll 1$, even the ``mobile'' fraction of the
electrons cannot carry DC~current
since their wave functions do not spread over macroscopic distances.
The doping has to exceed some ``percolation threshold'' to 
guarantee a finite DC~conductivity. 
Above the critical doping for the Anderson transition~$\delta_{\rm c}$
one should observe a metallic conductivity
at zero temperature but the magnetic signatures of the localized fraction
of the electrons (``local moments'') should remain. It requires further 
doping beyond~$\delta_{\rm c}$ to destroy the Curie behavior of 
the magnetic susceptibility.

Qualitatively similar results are found by
Dobrosavljevi\'c and Kotliar~\cite{Dobro}. 
They used the dynamical mean-field theory approach
and approximated the remaining local-impurity problem with the help
of the Slave Boson mean-field approach of
Barnes, Coleman, and Read and Newns~\cite{Barnes}--\cite{ReadNewns}.
They also address the strong-coupling limit and find that
local moments form in the metallic phase before the Anderson transition 
takes place. In their case the strength of the impurity potentials is varied
for fixed doping. In both cases the position of the mobility edge 
is varied with respect to the Fermi energy, 
and the physics should qualitatively be the same.
It should be clear, though, that our variational approach
gives only a rather crude description for the local moment formation
and the Anderson transition in the lower Hubbard band.

\section{Summary}
\label{summary}

In this work we introduced a general class of Gutzwiller-correlated
wave functions for multi-band Hubbard models. 
In contrast to earlier 
generalizations of the original Gutzwiller wave function~\cite{Gutzwiller2} 
to the case of degenerate 
bands~\cite{Gebhard2,ChaoGutz,Chao}
we introduced independent variational parameters 
for each multiple occupancy of a
lattice site. Only in this most general form 
each of the original variational parameters $g_{i;I}$ can be expressed by
the net occupancy densities $m_{i;I}$ which give the average probability
that configuration~$I$ is present on site~$i$.

We developed a diagrammatic
formalism which allows the approximation-free
evaluation of our general Gutzwiller-correlated 
wave functions in the limit of infinite spatial dimensions.
Our analytical results reproduce recent results within the Gutzwiller
approximation~\cite{Lu}--\cite{Buenemann}
and extend these to the case of a broken symmetry.
In addition, the Slave Boson mean-field theory~\cite{Dorin}--\cite{Fresard}
becomes variationally
controlled at zero temperature in the limit of infinite dimensions.
As can be inferred from the single-band case, the Slave Boson
approach at finite temperatures is limited to the Fermi-liquid
regime~\cite{Gebhard1,Gebhard2}; see also~\cite[Chap.~3.5]{BUCH}.
Lastly, we briefly discussed the results of the variational approach
to the Anderson transition in strongly correlated one-band systems.

In principle, we could go beyond the limitations
of the Gutzwiller approximation and systematically
calculate ground-state correlation functions and
$1/d$~corrections. However, the diagrammatic evaluation would be
very tedious, and we consider it unlikely that such an effort
will lead to qualitative changes of our understanding of
band-degenerate Hubbard models based on our variational approach.
For the moment it appears to be more rewarding to elucidate
the ground-state energy~(\ref{49}) in more detail. Thus far
this expression was investigated only for rather simple model 
systems~\cite{Okabe}--\cite{Hasegawa}.
These preliminary studies showed 
significant differences for the ground-state
magnetization as a function of the interaction strength
between the Hartree--Fock and the Gutzwiller-correlated
wave functions, as expected from the one-band case.
Furthermore, discontinuous metal-insulator transitions
occur for $N\geq 2$ for which the gap jumps to a finite value at the
critical interaction strength~\cite{Buenemann,Hasegawa}.

The investigation of the variational
ground-state phase diagram for realistic models, e.g., for five $d$~orbitals, 
is a numerically difficult task because 
the number of variational parameters exponentially increases 
with the number of orbitals.
Nevertheless, one may use various symmetries between
different multiple occupancies which considerably reduces
the number of independent variational parameters such that
the realistic case $N=5$ for transition metals and their compounds
should become tractable.

\ack

J.~B.\ thanks P.~Nozi\`eres for an invitation to
the ILL where this work was completed.


\section*{References}
\begin{thebibliography}{99}
%
\bibitem{Hubbard} Hubbard J 1963 \PRS~A~{\bf 276} 238
\item[] \dash\ 1964 \PRS~A~{\bf 281} 401
%
\bibitem{BUCH} Gebhard F 1997 {\it The Mott Metal--Insulator Transition -- Models and Methods},
Springer Tracts on Modern Physics~{\bf 137} (Berlin: Springer) 
%
\bibitem{Hubbard2} Hubbard J 1964 \PRS~A~{\bf 277} 237
%
\bibitem{Andersonimp} Anderson P W 1961 \PR~{\bf 124} 41
%
\bibitem{Varma} Varma C and Yafet Y 1976 \PR B~{\bf 13} 2950
%
\bibitem{Lu} Lu J P 1994 \PR~B~{\bf 49} 5687
\item[] \dash\ 1996 {\it Gutzwiller Approximation in Degenerate Hubbard Models}
(e-print cond-mat/9601133)
%
\bibitem{Okabe} Okabe T 1996 \JPSJ~{\bf 65} 1056
%
\bibitem{Buenemann} B\"unemann J and Weber W 1996 {\it The Generalized
Gutzwiller Method for $n\geq 2$ Correlated Orbitals: Itinerant Ferromagnetism
in d($e_g$)~Bands} (e-print cond-mat/9611031)
\item[] \dash\ 1997 \PR~B~{\bf 55} 4011
%
\bibitem{KR} Kotliar G and Ruckenstein A E 1986 \PRL~{\bf 57} 1362
%
\bibitem{Dorin} Dorin V and Schlottmann P 1995 \PR~B~{\bf 47} 5095
%
\bibitem{Hasegawa} Hasegawa H 1996 
{\it Slave-Boson Functional-Integral Approach
to the Hubbard Model with Orbital Degeneracy}
(e-print cond-mat/9612142)
\item[] \dash\ 1997 {\it Slave-Boson Mean-Field Theory of
the Antiferromagnetic State in the Doubly Degenerate Hubbard Model --
The Half-filled Case} (e-print cond-mat/9702034)
%
\bibitem{Fresard} Fr\'esard R and Kotliar G 1996
{\it Interplay of Mott Transition and Ferromagnetism in the Orbitally
Degenerate Hubbard Model} (e-print cond-mat/9612172)
%
\bibitem{Kajueter} Kajueter H and Kotliar G 1996 
{\it Band Degeneracy and Mott Transition: Dynamical Mean Field Study} 
(e-print cond-mat/9609176)
%
\bibitem{Rozenberg} Rozenberg M J 1997 \PR B {\bf 55} 4855
\item[] \dash\ {\it A Scenario for the Electronic State in the
Manganese Perovskites: The Orbital Correlated Metal}
(e-print cond-mat/9612089)
%
\bibitem{Gebhard1} Gebhard F 1990 \PR~B~{\bf 41} 9452 
%
\bibitem{Gebhard2} Gebhard F 1991 \PR~B~{\bf 44} 992
%
\bibitem{Revfour} Vollhardt D, van Dongen P G J, Gebhard F and Metzner W 1990
{\it Mod.~Phys.~Lett.}~B~{\bf 4} 499
%
\bibitem{Gunnarsson} Gunnarsson O, Koch E and Martin R M 1996
\PR~B~{\bf 54} 11026 
%
\bibitem{Gutzwiller2} Gutzwiller M C 1964 \PR~{\bf 134} A923 
%
\bibitem{ChaoGutz} Chao K A and Gutzwiller M C 1971 \JAP~{\bf 42} 1420
%
\bibitem{Chao} Chao K A 1971 \PR~B~{\bf 4} 4034
\item[] \dash\ 1973 \PR~B~{\bf 8} 1088
\item[] \dash\ 1974 \JPC~{\bf 7} 127
%
\bibitem{Fetter} Fetter A L and Walecka J D 1971 {\sl
Quantum Theory of Many-Particle Systems} (New York: McGraw--Hill)
%
\bibitem{MVdinfty} Metzner W and Vollhardt D 1989 \PRL~{\bf 62} 324
%
\bibitem{Vollrev} Vollhardt D 1984 \RMP~{\bf 56} 99
%
\bibitem{Gutzi63} Gutzwiller M C 1963 \PRL~{\bf 10} 159
\item[] \dash\ 1965 \PR~{\bf 137} A1726
%
\bibitem{Metzner} Metzner W and Vollhardt D 1988 \PR B {\bf 37} 7382
%
\bibitem{Ogawa} Ogawa T, Kanda K and Matsubara T 1975 
{\it Prog.~Theor.~Phys.}~{\bf 53} 614
%
\bibitem{Yoko} Yokoyama H and Shiba H 1990 \JPSJ~{\bf 59} 3669
%
\bibitem{Michael} Dzierzawa M and Fr\'{e}sard R 1993 \ZP~B~{\bf 91}
245
%
\bibitem{Logan} Logan D E, Szczech Y H and Tusch M E 1995
{\it Metal--Insulator Transitions Revisited}
ed P P Edwards and C R N Rao (London: Taylor and Francis)
p.~395
%
\bibitem{Barnes} Barnes J 1977 \JPF~{\bf 7} 2637
%
\bibitem{Coleman} Coleman P 1983 \PR B~{\bf 28} 5255
%
\bibitem{ReadNewns} Read N and Newns D M 1983 \JPC~{\bf 16} 3273
%
\bibitem{Dobro} Dobrosavljevi\'c V and Kotliar G 1996 {\it
Mean Field Theory of the Mott--Anderson Transition}
(e-print cond-mat/9611100)
%
\end{thebibliography}
\end{document}